\documentclass{pasa}%

\title[The Role of Discs in Prestellar Cores]{The Role of Discs in the Collapse and Fragmentation of Prestellar Cores}
\author[O. Lomax, A. P. Whitworth, D. A. Hubber]{O. Lomax\thanks{E-mail: oliver.lomax@astro.cf.ac.uk}$^1$, A. P. Whitworth$^1$, D. A. Hubber$^{2,3}$\\
\affil{$^1$School of Physics and Astronomy, Cardiff University, Cardiff CF24 3AA, UK}%
\affil{$^2$University Observatory, Ludwig-Maximilians-University Munich, Scheinerstr.1, D-81679 Munich, Germany}
\affil{$^3$Excellence Cluster Universe, Boltzmannstr. 2, D-85748 Garching, Germany}}%
\jid{PASA}
\doi{10.1017/pas.\the\year.xxx}
\jyear{\the\year}

\usepackage[authoryear]{natbib}
\bibpunct{(}{)}{;}{a}{}{,}
\usepackage{amsmath,upgreek,subfigure}

\newcommand{\changes}[1]{{#1}}

\begin{document}

\begin{abstract}

Disc fragmentation provides an important mechanism for producing low mass stars in prestellar cores. Here, we describe Smoothed Particle Hydrodynamics simulations which show how populations of prestellar cores evolve into stars. We find the observed masses and multiplicities of stars can be recovered under certain conditions.

First, protostellar feedback from a star must be episodic. The continuous accretion of disc material on to a central protostar results in local temperatures which are too high for disc fragmentation. If, however, the accretion occurs in intense outbursts, separated by a downtime of $\sim10^4\,\mathrm{years}$, gravitational instabilities can develop and the disc can fragment.

Second, a significant amount of the cores' internal kinetic energy should be in solenoidal turbulent modes. Cores with less than a third of their kinetic energy in solenoidal modes have insufficient angular momentum to form fragmenting discs. In the absence of discs, cores can fragment but results in a top heavy distribution of masses with very few low mass objects.

\end{abstract}

\maketitle

\section{INTRODUCTION}


Two of the main goals of star formation theory are (i) to understand the origin of the stellar initial mass function (IMF) \citep[e.g.][]{K01,C03,C05} and (ii) to explain the properties of stellar multiple systems \citep[e.g.][]{RMH10,JHB12}. One possible solution to this problem is the turbulent fragmentation of giant molecular clouds. Here, turbulent flows within molecular clouds produce dense cores of gas \citep[e.g.][]{PN02,HC08,HC09} of order $0.01\text{ to }0.1\,\mathrm{pc}$ across. These may be Jeans unstable, in which case they collapse to form stars \citep[e.g.][]{AWB93,AWB00}. This has been demonstrated in numerical simulations by \citet{B98,B00,HBB01,MH03, GW04,DCB04a,DCB04b,GWW04,GWW06,WBWNG09,WNWB10,WWG12,LWHSW14,LWHSW14b,LWH15}.

Observations of prestellar cores \citep[e.g.][]{MAN98,TS98,JWM00,MAWB01,JFMM01,SSGK06,EYG06,JB06,NW-T07,ALL07,EES08,SNW-T08,RLM09,KAM10,PWK15} show that the core mass function (CMF) is very similar in shape to the IMF (i.e. a lognormal distribution with a power-law tail at high mass), albeit shifted upwards in mass by a factor of three to five. This had lead to the suggestion that there is a self-similar mapping of the CMF onto the IMF. Statistical analysis by \citet{HWGW13} suggests that a core should spawn on average four to five stars in order to explain the observed abundance of multiple systems.

A core may fragment into multiple objects via either turbulent fragmentation -- similar to the molecular cloud -- or disc fragmentation. Observed $\mathrm{N_2H^+}$ line widths in cores indicate that the internal velocity dispersion in most cores is sub to trans-sonic. This suggests that a typical core is unlikely to collapse and fragment into more than one or two objects through turbulence alone. However, the first protostars to form are usually attended by accretion discs \citep{KH95}. These discs may fragment if two criteria are fulfilled. First, the disc must have a sufficiently large surface density $\varSigma(R)$ so that fragments can overcome thermal and centrifugal support,
\begin{equation}
  \varSigma(R)\gtrsim\frac{c(R)\,\kappa(R)}{\uppi\,G},
  \label{eqn:tcrit}
\end{equation}
where $\kappa(R)$ is the epicyclic frequency and $c(R)$ is the sound speed; \citep{T64}. Second, the cooling time of a fragment must be shorter than the orbital period if it is to avoid being sheared part \citep{Gam01}.

\changes{The above criteria apply well to low-mass discs. However, disc dynamics are more complex when the disc and the central protostar have similar masses. In cases where the disc is marginally unstable (i.e. $t_\textsc{cool}\sim t_\textsc{orbit}$ and $\uppi\,G\,\varSigma\sim c\,\kappa$), instabilities develop which bolster the accretion rate onto the central protostar. This lowers the mass (and hence surface density) of the disc, restabilising it. This suggests that an otherwise unstable disc may be able to remain stable by undergoing episodic accretion events \citep{LR05}.}

\changes{When non-local effects such as radiative transfer dominate the disc temperature, local cooling timescales become very difficult -- if not impossible -- to calculate. Here, again, a seemingly unstable disc may be able to resist fragmentation \citep[e.g.][]{TTMI15}. \citet{FR13} also show that irradiation (from a central protostar and/or interstellar radiation field) increases the jeans mass within disc spiral structures. They deduce that disc fragmentation is far more likely to result in low mass stars and brown dwarfs than gas giant planets.}

In this paper, we analyse simulations which follow the evolution of an ensemble of synthetic cores, based on the properties of those in the Ophiuchus star forming region \citep[See][for more details.]{LWHSW14,LWHSW14b,LWH15}. We examine how protostellar feedback affects the fragmentation of discs, and how the results compare with observed stars. We also examine how the the ratio of solenoidal to compressive modes in the turbulent velocity field affects the formation of discs and filaments within a prestellar core. In Section \ref{sec:method} we describe the numberical method used to evolve the cores. In Section \ref{sec:cores} we describe (i) how to generate realistic core initial conditions and (ii) how they evolve under different prescriptions of radiative feedback. In Section \ref{sec:turb} we show how changing the ratio of solenoidal to compressive modes in the velocity field affects core fragmentation. We summarise and conclude in Section \ref{sec:summary}\,.

\section{NUMERICAL METHOD}
\label{sec:method}

\subsection{Smoothed particle hydrodynamics}

Core evolution is simulated using the \textsc{seren} $\nabla h$-SPH code \citep{HBMW11}. Gravitational forces are computed using a tree and artificial viscosity is controlled by the \citet{MM97} prescription. In all simulations, the SPH particles have mass $m_\textsc{sph}=10^{-5}\,\mathrm{M}_{\odot}$, so that the opacity limit ($\sim\!\!3\times10^{-3}\,\mathrm{M}_{\odot}$) is resolved with $\sim\!\!300$ particles. Gravitationally bound regions with density higher than $\rho_\textsc{sink}=10^{-9}\,\mathrm{g}\,\mathrm{cm}^{-3}$ are replaced with sink particles \citep{HWW13}. Sink particles have radius $r_\textsc{sink}\simeq0.2\,\mathrm{au}$, corresponding to the smoothing length of an SPH particle with density equal to $\rho_\textsc{sink}$. The equation of state and the energy equation are treated with the algorithm described in \citet{SWBG07}. Magenetic fields and mechanical feedback (e.g. stellar winds) is not included in these simulations.

\subsection{Accretion Feedback}

Radiative feedback from the protostars (i.e. sink particles formed in the simulations) is included. The dominant contribution to the luminosity of a protostar is usually from accretion,
\begin{eqnarray}
   L_\star&\simeq&\frac{f\,G\,M_\star\,\dot{M}_\star}{R_\star}\,,
   \label{star_lum}
\end{eqnarray}
where, $f=0.75$ is the fraction of the accreted material's gravitational energy that is radiated from the surface of the protostar \citep[the rest is presumed to be removed by bipolar jets and outflows; ][]{OKMK09}, $M_\star$ is the mass of the protostar, $\dot{M}_\star$ is the rate of accretion onto the protostar and $R_\star=3\,\mathrm{R}_{\odot}$ is the approximate radius of a protostar \citep{PS93}.

We adopt the phenomenological model of episodic accretion, presented by \citet{SWH11,SWH12}. This is based on calculations of the magneto-rotational instability (MRI) \citep{ZHG09,ZHG10,Z10}. In the outer disc of a protostar (outside the sink radius, and therefore resolved by the simulation), angular momentum is redistributed by gravitational torques and material spirals inwards towards the sink. At distances within the sink radius (unresolved by the simulation), the inner disc is so hot that it is gravitationally stable, and unable to fragment. Material continues to accreted onto this inner disc (i.e. the sink particle) until the gas is hot enough to thermally ionise and couple with the local magnetic field. The MRI cuts in and magnetic torques allow material in the inner disc to rapidly accrete onto the protostar. This results in extended periods of very low accretion luminosity, punctuated by intense, episodic outbursts. The length of the downtime between outbursts -- during which disc fragmentation may occur -- is given by,
\begin{equation}
  \begin{split}
  t_\textsc{down}&\sim1.3\times10^4\,\mathrm{years}\\
  &\times\left(\frac{M_\star}{0.2\,\mathrm{M}_{\odot}}\right)^{2/3}\left(\frac{\dot{M}_\textsc{sink}}{10^{-5}\,\mathrm{M}_{\odot}\mathrm{yr}^{-1}}\right)^{-8/9}\,,
  \label{eqn:downtime}
  \end{split}
\end{equation}
where $\dot{M}_\textsc{sink}$ is the rate at which material flows into the sink.

There is also observational motivation for adopting an episodic model. The luminosities of young stars are about an order of magnitude lower than expected from continuous accretion \citep[this is the \emph{luminosity problem}, first noted by][]{KHSS90}. Furthermore, FU Ori-type stars \citep[e.g.][]{H77,GAR08,PSMB10,G11,PKG13} can exhibit large increases in luminosity which last $\lesssim10^2\,\mathrm{years}$. Statistical arguments by \citet{SFW13} suggest that the downtime between outbursts should be of order $10^4\,\mathrm{years}$, similar to the timescale given in Eqn. \ref{eqn:downtime}\,.

\section{PRESTELLAR CORES IN OPHIUCHUS}
\label{sec:cores}

\subsection{Initial conditions}

Using the observed properties of cores as a basis for numerical simulations presents a difficult inverse problem. The mass, temperature, projected area and projected aspect ratio of a core can be reasonably inferred from bolometric measurements of the dust in prestellar cores. In addition, the line-of-sight velocity dispersion within the core can be inferred from the width of molecular lines. However, the initial boundary conditions of a core simulation must represent the full spatial and velocity structure of the system. This occupies six dimensions, whereas observational data can only provide information on three (i.e. two spatial and one velocity).

Rather that trying to emulate a specific core -- which is arguably an impossibility -- we can relatively easily define a distribution of cores which have the same, or at least very similar, statistical properties to those in a given region. We based the synthetic core initial conditions on Ophiuchus. This is a well studied region, for which many of the aforementioned core properties have been measured.

\subsubsection{Mass, size and velocity dispersion}

Only some of the measured core masses in Ophiuchus have both an associated size and velocity dispersion. In order to make the most of the data, We define the following lognormal probability distribution of $\boldsymbol{x}\equiv(\log(M),\log(R),\log(\sigma_\textsc{nt}))$:
\begin{equation}
   P(\boldsymbol{x})=\frac{1}{(2\uppi)^{3/2}|\boldsymbol{\varSigma}|}\exp\left(-\frac{1}{2}(\boldsymbol{x}-\boldsymbol{\mu})^\mathrm{T}\boldsymbol{\varSigma}^{-1}(\boldsymbol{x}-\boldsymbol{\mu}) \right)\,,
   \label{obsdist}
\end{equation}
where
\begin{equation}
   \boldsymbol{\mu}\equiv
      \begin{pmatrix}
         \mu_{_M} \\
         \mu_{_R} \\
         \mu_{_{\sigma_\textsc{nt}}}
      \end{pmatrix}\,,
\end{equation}
and
\begin{equation}
      \boldsymbol{\varSigma}\equiv
      \begin{pmatrix}
         \sigma_{_M}^2 & \rho_{_{M,R}}\,\sigma_{_M}\sigma_{_R} & \rho_{_{M,\sigma_\textsc{nt}}}\,\sigma_{_M}\sigma_{_{\sigma_\textsc{nt}}} \\
         \rho_{_{M,R}}\,\sigma_{_M}\sigma_{_R} & \sigma_{_R}^2 & \rho_{_{R,\sigma_\textsc{nt}}}\,\sigma_{_R}\sigma_{_{\sigma_\textsc{nt}}} \\
         \rho_{_{M,\sigma_\textsc{nt}}}\,\sigma_{_M}\sigma_{_{\sigma_\textsc{nt}}} & \rho_{_{R,\sigma_\textsc{nt}}}\,\sigma_{_R}\sigma_{_{\sigma_\textsc{nt}}} & \sigma_{_{\sigma_\textsc{nt}}}^2
      \end{pmatrix}\,.
\end{equation}
The coefficients of $\boldsymbol{\mu}$ and $\boldsymbol{\varSigma}$ are calculated from the observed Ophiuchus data \citep{MAN98,ABMP07} and are given in Table \ref{log_params}\,. From $P(\boldsymbol{x})$, We are able to draw any number of masses, sizes and velocity dispersions, all of which were statistically similar to those in Ophiuchus.

\begin{table}
   \centering
   \begin{tabular}{llc}\hline
   Parameter & & Value \\\hline
   $\mu_{_M}$ & $[\log(M/\mathrm{M_{\odot}})]$ & -0.57 \\
   $\mu_{_R}$ & $[\log(R/\mathrm{AU})]$ & 3.11 \\
   $\mu_{_{\sigma_\textsc{nt}}}$ & $[\log(\sigma_\textsc{nt}/\mathrm{km\,s}^{-1})]$ & -0.95 \\
   $\sigma_{_M}$ & $[\log(M/\mathrm{M_{\odot}})]$ & 0.43 \\
   $\sigma_{_R}$ & $[\log(R/\mathrm{AU})]$ & 0.27 \\
   $\sigma_{_{\sigma_\textsc{nt}}}$ & $[\log(\sigma_\textsc{nt}/\mathrm{km\,s}^{-1})]$ & 0.20 \\
   $\rho_{_{M,R}}$ && 0.61 \\
   $\rho_{_{M,\sigma_\textsc{nt}}}$ && 0.49 \\
   $\rho_{_{R,\sigma_\textsc{nt}}}$ && 0.11 \\\hline
   \end{tabular}
   \caption{Arithmetic means, standard deviations and correlation coefficients of $\log(M)$, $\log(R)$ and $\log(\sigma_\textsc{nt})$ for cores in Ophiuchus.}
   \label{log_params}
\end{table}

\begin{figure*}
   \centering
   \includegraphics[width=\textwidth]{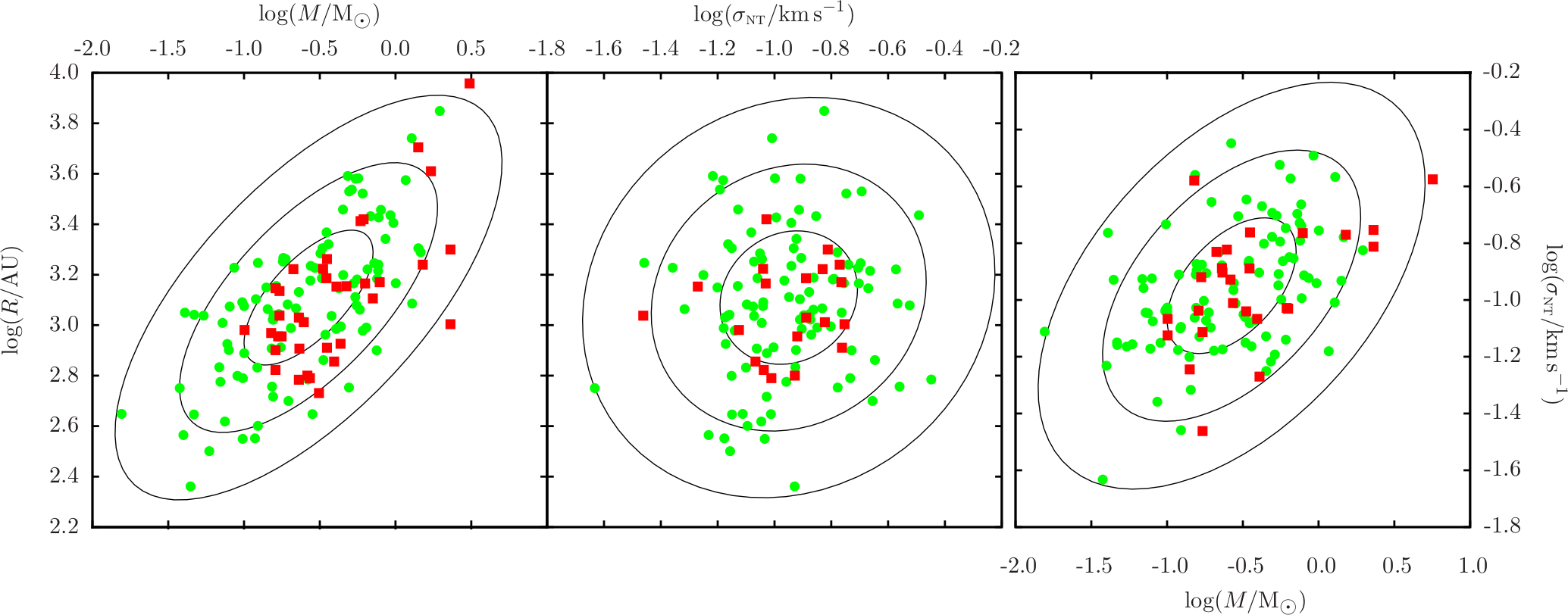}
   \caption[Lognormal distribution of core parameters.]{The multivariate lognormal distribution, $P(\boldsymbol{x})$ where \mbox{$\boldsymbol{x}=(\log(M),\log(R),\log(\sigma_\textsc{nt}))$}. The figure shows the projections through $\log(\sigma_\textsc{nt})$, through $\log(M)$ and through $\log(R)$. The concentric ellipses show the $1\sigma$, $2\sigma$ and $3\sigma$ regions of the distribution. The green circles are randomly drawn points from $P(\boldsymbol{x})$. The red squares are the observational data from \citet{MAN98} and \citet{ABMP07}.\changes{ See \citet{LWHSW14} for the orginal version of this figure.}}
   \label{core_dist}
\end{figure*}

\subsubsection{Shape}

Molecular cloud cores often have elongated, irregular shapes. We include this in the simulations by assuming that each intrinsic core shape can be drawn from a family of triaxial ellipsoids. Each ellipsoid had axes:
\begin{equation}
  \begin{split}
    A&=1\,,\\
    B&=\exp(\tau\mathcal{G}_\textsc{b})\,,\\
    C&=\exp(\tau\mathcal{G}_\textsc{c})\,,
  \end{split}
\label{EQN:M1}
\end{equation}
where $\mathcal{G}_\textsc{b}$ and $\mathcal{G}_\textsc{c}$ are random numbers drawn from a Gaussian distribution with zero mean and unit standard deviation. The scale-parameter $\tau\approx0.6$ is a fit to the distribution of projected aspect ratios in Ophiuchus \citep{LWC13}. The axes are normalised to a given $R$, giving the dimensions of the core,
\begin{equation}
   \begin{split}
      A_\textsc{core}=&\frac{R}{(BC)^{\frac{1}{3}}}\,,\\
      B_\textsc{core}=&BA_\textsc{core}\,,\\
      C_\textsc{core}=&CA_\textsc{core}\,.
   \end{split}
\end{equation}

\subsubsection{Density profile}

Density profiles of cores are often well fitted by those of critical Bonnor-Ebert spheres \citep[e.g.][]{B56,ALL01,HWL01,LMR08}. We use such a profile for the ellipsoidal cores. Here, $\rho=\rho_{_{\rm C}}{\rm e}^{-\psi(\xi)}$, where $\rho_{_{\rm C}}$ is the central density, $\psi$ is the Isothermal Function, $\xi$ is the dimensionless radius, $\xi_{_{\rm B}}=6.451$ is the boundary of the sphere. The density at any given point $(x,y,z)$, where $(0,0,0)$ is the centre of the core, is given by
\begin{eqnarray}
\xi&=&\xi_{_{\rm FWHM}}\left(\frac{x^2}{A_\textsc{core}^2}+\frac{y^2}{B_\textsc{core}^2}+\frac{z^2}{C_\textsc{core}^2}\right)^{1/2}\,,\\
\rho(\xi)&=&\frac{M_\textsc{core}\xi_{_{\rm B}}{\rm e}^{-\psi(\xi)}}{4\pi A_\textsc{core}B_\textsc{core}C_\textsc{core}\psi'(\xi_{_{\rm B}})}\,,\hspace{0.5cm}\xi<\xi_{_{\rm B}}\,,
\end{eqnarray}
where $\psi'$ is the first derivative of $\psi$ and $\xi_{_{\rm FWHM}}=2.424$ is the Full Width at Half Maximum of the column density through a critical Bonnor-Ebert sphere.

\subsubsection{Velocity field}

Each core is given a turbulent velocity field with power spectrum $P\propto k^{-4}$ in three dimensions. We include bulk rotation and radial excursion by modifying the amplitudes $\boldsymbol{a}$ and phases $\boldsymbol{\varphi}$ of the $k=1$ modes:
\begin{equation}
      \begin{split}
         \begin{bmatrix}
            \boldsymbol{a}(1,0,0) \\
            \boldsymbol{a}(0,1,0) \\
            \boldsymbol{a}(0,0,1)
         \end{bmatrix}&=
         \begin{bmatrix}
            r_x & \omega_z & -\omega_y \\
            -\omega_z & r_y & \omega_x \\
            \omega_y & -\omega_x & r_z
         \end{bmatrix}\,,\\
         \begin{bmatrix}
            \boldsymbol{\varphi}(1,0,0) \\
            \boldsymbol{\varphi}(0,1,0) \\
            \boldsymbol{\varphi}(0,0,1)
         \end{bmatrix}&=
         \begin{bmatrix}
            \uppi/2 & \uppi/2 & \uppi/2 \\
            \uppi/2 & \uppi/2 & \uppi/2 \\
            \uppi/2 & \uppi/2 & \uppi/2
         \end{bmatrix}\,.
      \end{split}
\end{equation}
The amplitude components $r_x$, $r_y$, $r_z$, $\omega_x$, $\omega_y$ and $\omega_z$ are drawn independently from a Gaussian distribution with zero mean and unit variance. The $r$ terms define the amount of excursion along a given axis. The $\omega$ terms define the amount rotation about a given axis. The fields are generated on a $128^3$ grid and interpolated onto the SPH particles. The velocity dispersion of the particles is normalised to a given value of $\sigma_\textsc{nt}$.

\subsection{Results}

One hundred synthetic cores have been evolved for $2\times10^5\,\mathrm{years}$. This is roughly an order of magnitude greater than the average core free-fall time and of the same order as the estimated core-core collisional timescale \citep{ABMP07}. Of the one hundred cores, sixty are prestellar. Each simulation has been performed three times: once with no radiative feedback from accretion (NRF); again with episodic feedback from accretion (ERF); and finally with continuous feedback from accretion (CRF). With CRF, the protostellar luminosity was calculated using Eqn. \ref{star_lum}, with $\dot{M}_\star$ set to $\dot{M}_\textsc{sink}$.

\subsubsection{Stellar masses}
\label{sec:masses}

\begin{figure}
\centering
\includegraphics[width=\columnwidth]{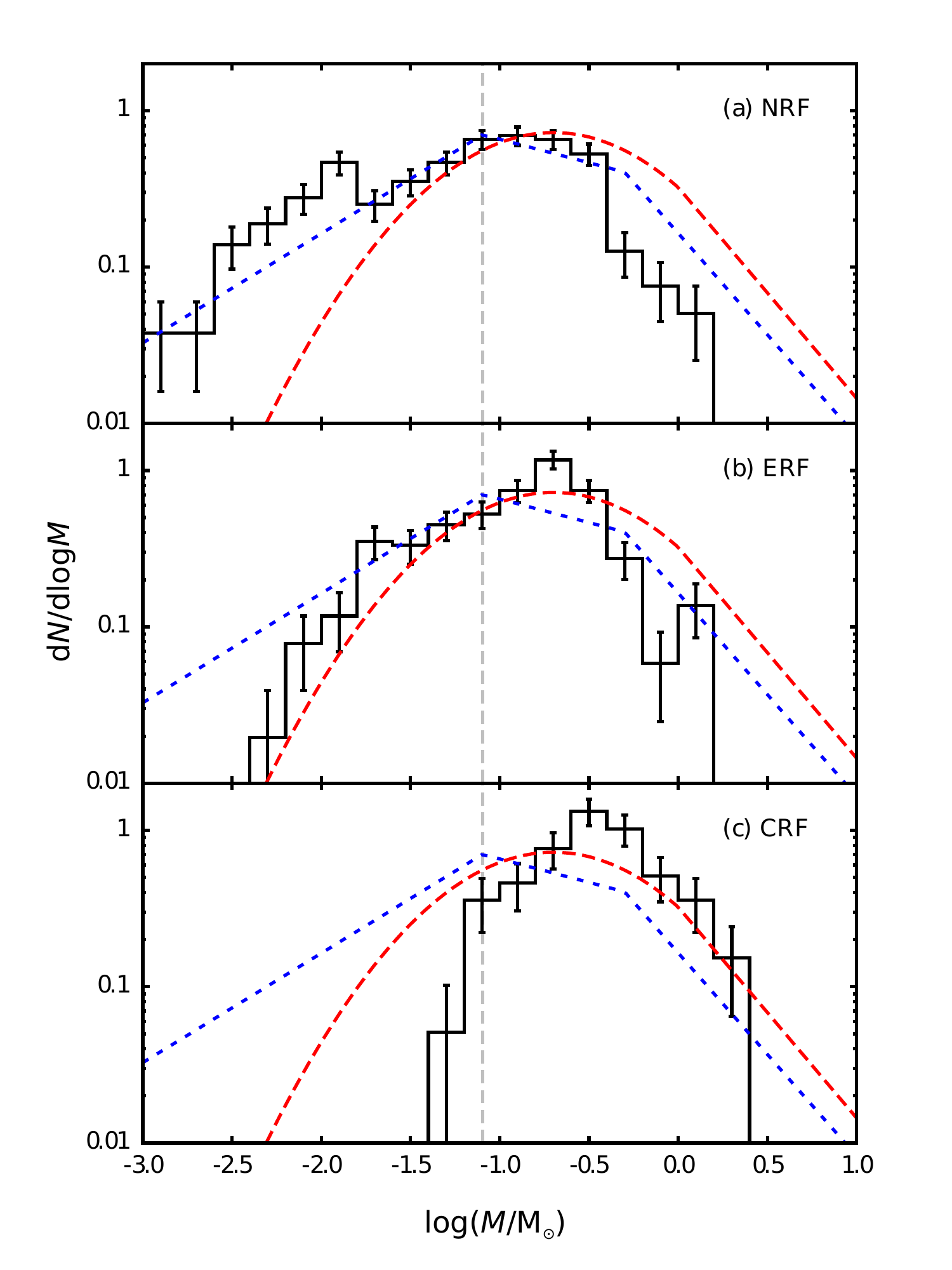}
\caption{The black histograms show stellar mass functions for (a) NRF, (b) ERF, and (c) CRF. The blue dotted straight lines, and the red dashed lognormal curve, show, respectively, the \citet{C05} and \citet{K01} fits to the observed IMF. The vertical dashed line shows the hydrogen burning limit at $M=0.08\,\mathrm{M_\odot}$.\changes{ See \citet{LWHSW14b} for the orginal version of this figure.}}
\label{masses1}
\end{figure}

We find two very noticeable trends when progressing along the series NRF$\to$ERF$\to$CRF. First, the median number stars formed per core decreases: $\mathcal{N}_\textsc{s/c}^\textsc{nrf}=6.0^{+4.0}_{-2.0}$, $\mathcal{N}_\textsc{s/c}^\textsc{erf}=3.5^{+3.5}_{-2.5}$, $\mathcal{N}_\textsc{s/c}^\textsc{crf}=1.0^{+0.0}_{-0.0}$. Second, the median protostellar mass (see Fig. \ref{masses1}) shifts upwards: $\log_{10} M_\star^\textsc{nrf}=-1.1^{+0.4}_{-0.6}$, $\log_{10} M_\star^\textsc{erf}=-0.8^{+0.2}_{-0.4}$, $\log_{10} M_\star^\textsc{crf}=-0.5^{+0.2}_{-0.2}$. \footnote{The uncertainties give the interquartile range of the distribution.}

These trends occur because disc fragmentation is strongly affected by protostellar feedback. In the case with NRF, discs are relatively cold and easily fragment. Recall from Eqn. \ref{eqn:tcrit} that fragmentation can occur if the disc has a sufficiently large column density \emph{or} the sound speed is sufficiently low. With ERF, fragmentation is occasionally interrupted by the episodic outbursts. With CRF, the discs are constantly heated, and fragmentation becomes difficult. As a consequence, the central protostar or binary usually accretes the entire mass of its disc. Protostars with NRF, and to a slightly lesser extent ERF, are often attended by multiple low mass companions which partially starve the primary protostar of the remaining gas. Of the three sets of simulations, those with ERF best reproduce the \citet{C05} IMF.

In star forming regions, the observed ratio of stars to brown dwarfs is
\begin{equation}
  \mathcal{A}=\frac{N(0.08\,\mathrm{M}_{\odot}<M\leq1.0\,\mathrm{M}_{\odot})}{N(0.03\,\mathrm{M}_{\odot}<M\leq0.08\,\mathrm{M}_{\odot})}=4.3\pm1.6\,,
\end{equation}
\citep[See][]{AMGA08}. This figure is best reproduced with ERF: $\mathcal{A}_\textsc{erf}=3.9\pm0.6$.\footnote{Here the uncertainty is calculated from the Poison counting error.} Simulations with NRF and CRF yield $\mathcal{A}_\textsc{nrf}=2.2\pm0.3$ and $\mathcal{A}_\textsc{crf}=17\pm8$. While the figure for NRF is not completely incompatible with observations, the value with CRF is far too high. This is because brown dwarfs are unable to form via disc fragmentation \citep[e.g.][]{SHW07,SW09a}.

\subsubsection{Stellar multiplicities}
\label{sec:multiplicity}

The core simulations also produce a wide variety of multiple systems. These are either simple binary systems or high order ($N\ge3$) hierarchical multiples. A high order system can be thought of as a binary where one or both components is another binary system. Multiple systems of protostars are extracted from the end state simulations if they have been tidally stable for at least one orbital period.

\begin{figure}
\centering
\includegraphics[width=\columnwidth]{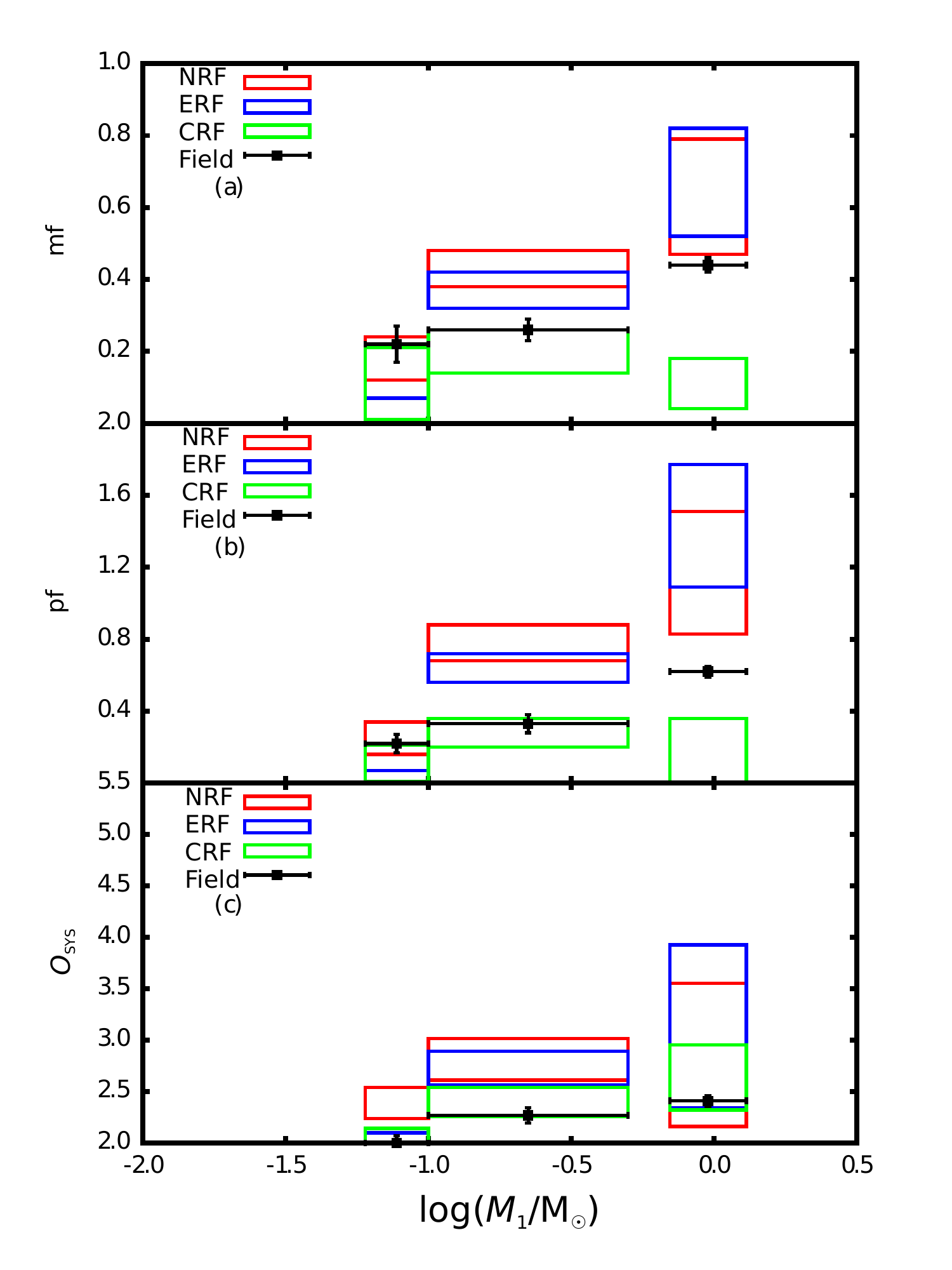}
\caption{Multiplicity frequency (a), pairing factor (b) and mean system order (c) for systems with very low mass, M-dwarf and solar type primaries. The red boxes give the values for the NRF simulations, blue for the ERF and green for the CRF. The black points give the values observed in field Main Sequence stars. In all cases, the width of a box shows the extent of the mass bin, and the height shows the uncertainty.}
\label{fig:multfreq}
\end{figure}

There are many ways to statistically describe the multiplicity of a population of stellar systems. Here, following \citet{RZ93}, we use the \emph{multiplicity frequency} and the \emph{pairing factor}. The multiplicity frequency is fraction of systems which is multiple,
\begin{equation}
  \mathrm{mf}=\frac{B+T+Q\dots}{S+B+T+Q\dots}\,,
\end{equation}
where $S$ is the number of single systems, $B$ is the number of binaries, $T$ is the number of triples, etc.. The pairing factor is the average number of orbits per system,
\begin{equation}
  \mathrm{pf}=\frac{B+2T+3Q\dots}{S+B+T+Q\dots}\,.
\end{equation}
These two quantities are particularly useful in conjunction, as their ratio gives the average number of objects per multiple system,
\begin{equation}
  \mathcal{O}_\textsc{sys}=1+\frac{\mathrm{pf}}{\mathrm{mf}}\,
\end{equation}
By definition $\mathcal{O}_\textsc{sys}\ge2$\,.

\begin{figure*}
  \centering
  \subfigure[$t=2.2\times10^4\,\mathrm{yrs}$]{\label{sex_1}\includegraphics[width=0.45\textwidth]{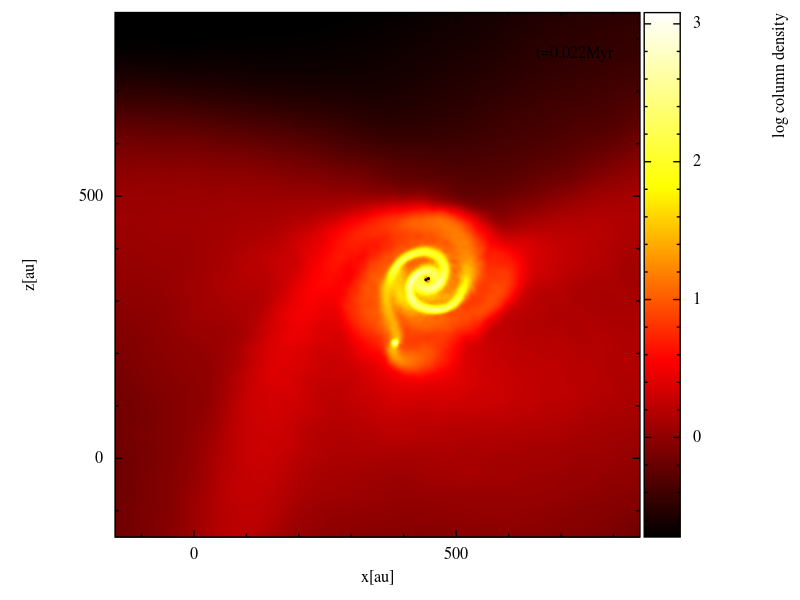}}
  \subfigure[$t=2.7\times10^4\,\mathrm{yrs}$]{\label{sex_2}\includegraphics[width=0.45\textwidth]{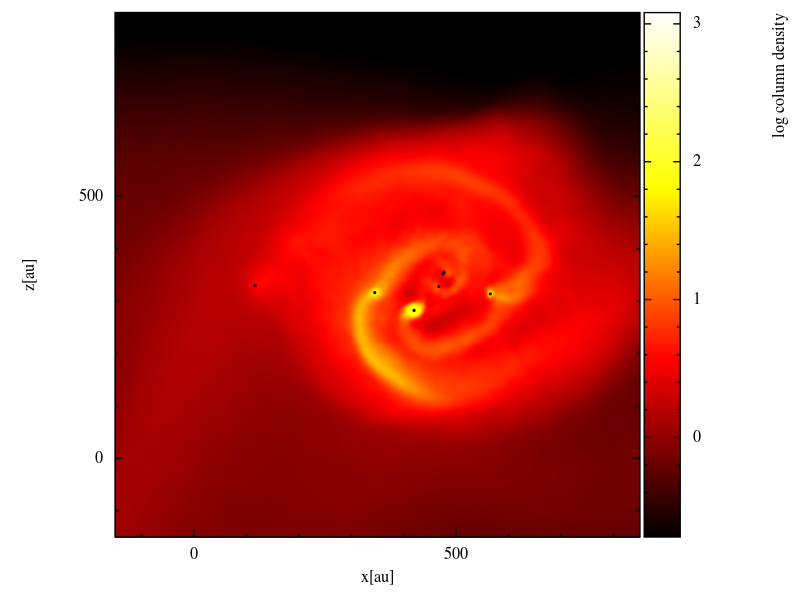}}
  \subfigure[$t=3.2\times10^4\,\mathrm{yrs}$]{\label{sex_3}\includegraphics[width=0.45\textwidth]{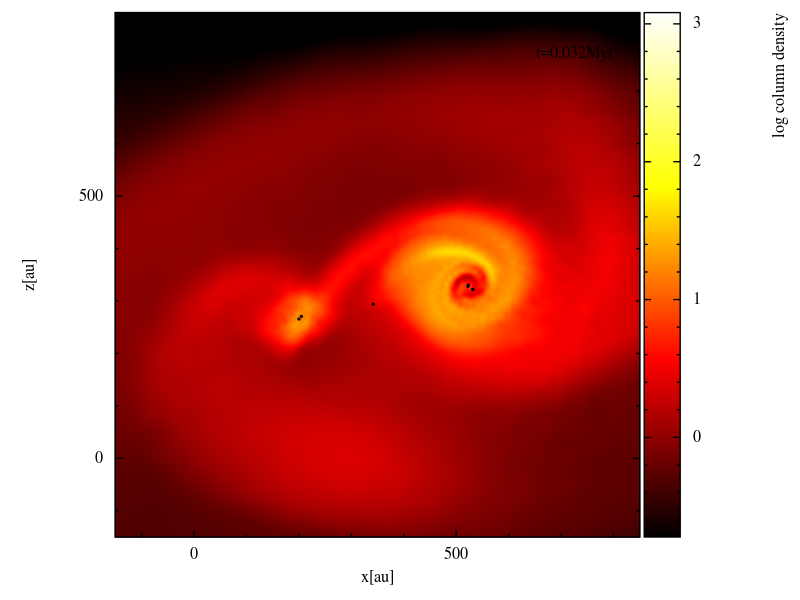}}
  \subfigure[$t=3.7\times10^4\,\mathrm{yrs}$]{\label{sex_4}\includegraphics[width=0.45\textwidth]{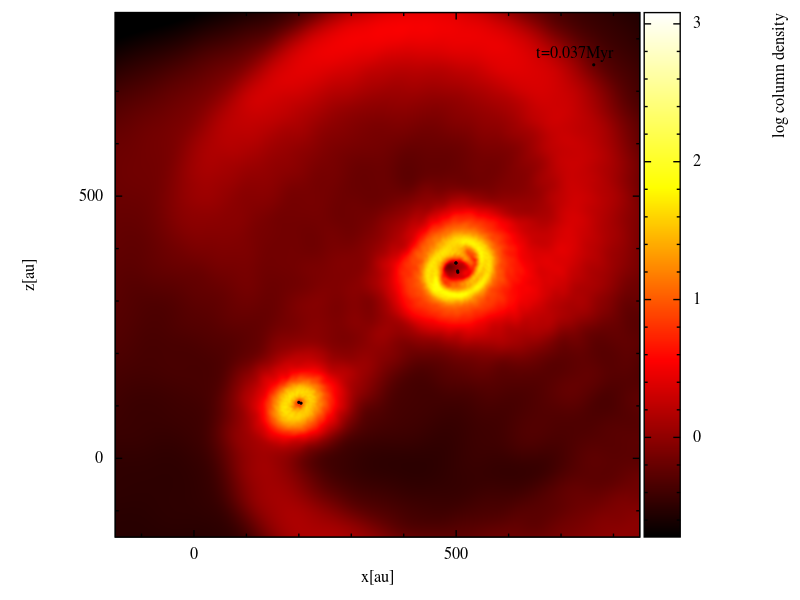}}
  \caption{A sequence of column density maps of a core during disc fragmentation. The initial core has $M=1.3\,\mathrm{M_\odot}$, $R=3000\,\mathrm{au}$, $\sigma_\textsc{nt}=0.3\,\mathrm{km\,s^{-1}}$ and is evolved with ERF. The colour-bar gives shows the column density in units of $\mathrm{g\,cm^{-2}}$. The black dots show the positions of sink particles, i.e. protostars. Fig. \ref{sex_1} shows gravitational instabilities developing in a circumbinary disc. Fig. \ref{sex_2} shows seven protostars in an unstable configuration. In Fig. \ref{sex_3}, the protostars are configured in a quadruple system (right) and a binary (left); a single protostar is being ejected (centre). Fig. \ref{sex_4} shows a stable sextuple system which lasts until the end of the simulation.}
  \label{fig:sextuple}
\end{figure*}

Fig. \ref{fig:multfreq} shows mf, pf and $\mathcal{O}_\textsc{sys}$ for systems with very low mass or brown dwarf primaries, M-dwarf primaries and solar-type primaries\footnote{Here we define very low mass stars and brown dwarfs as stars with $0.06\,\mathrm{M_\odot}\leq M<0.1\,\mathrm{M_\odot}$, M-dwarfs as stars with $0.1\,\mathrm{M_\odot}\leq M<0.5\,\mathrm{M_\odot}$ and solar types as stars with $0.7\,\mathrm{M_\odot}\leq M<1.3\,\mathrm{M_\odot}$.}. For comparison, we have also included the same figure for stars in the field \citep[See][and references therein]{DK13}. Observations of pre-Main Sequence stars in Ophiuchus and Taurus suggest that multiplicity is high when stars are young \citep[e.g.][]{LZW93,RKL05,KIM11}. As stars age, particularly if they are in a clustered environment, dynamical interactions erode these systems and the multiplicity frequency drops to that observed in the field \citep[e.g.][]{K95a,PGKK09,PG11,PG12}. The multiplicity of protostars (i.e. the objects formed in these simulations) should therefore be higher than the multiplicity of field stars. This requirement is met for the simulations with NRF and ERF. However, with CRF, the multiplicity frequency is too low for solar-type primaries.

Very high order systems, e.g. sextuples, are also form in these simulations. Fig. \ref{fig:sextuple} shows a sextuple system formed from a solar mass core with ERF. Initially the core fragments into two objects, which form a binary system with a circumbinary disc. This disc fragments into a further five objects. It finally settles into a sextuple system (a binary system orbiting a quadruple) with separations ranging from $\sim600\,\mathrm{au}$ for the outer orbit to $\sim0.1\,\mathrm{au}$ for the inner-most orbits. Sextuple systems similar to this are found in star forming regions \citep{KIM11} and the field \citet{ET08,T08}.


All three sets of simulations produce systems with semi-major axes ranging between roughly $0.1\,\mathrm{au}$ (the resolution limit of the simulation) and $1000\,\mathrm{au}$. This upper limit corresponds to that observed in star forming regions \citep[e.g.][]{KPPG12,KGPP12}. Systems with wider orbits ($\gtrsim10^3\,\mathrm{au}$) are probably assembled later through dynamical interactions in clustered environments \citep[e.g.][]{VeryWideBinaries}.

\subsection{Summary}

\changes{We find that the mass distribution and multiplicity statistics of young stars can be recovered from simulations if radiative feedback from protostellar accretion is episodic}. The periods of low luminosity provide a window of opportunity, during which protostellar discs can fragment. When feedback is continuous, the disc is too warm to permit disc fragmentation. In this instance, the simulations result in (i) a top-heavy IMF, and (ii) too few stars per core to satisfactorily reproduce observed multiplicity statistics. When there is no radiative feedback, good multiplicity statistics are recovered, but the resultant IMF has too many brown dwarfs.

\section{TURBULENCE: DISCS AND FILAMENTS}
\label{sec:turb}

To examine how the structure of a core's velocity field affects the star formation process, we take a single core set-up and vary the structure of the velocity field. The core is spherical, with $M=3\,\mathrm{M_\odot}$, $R=3000\,\mathrm{au}$ and $\sigma_\textsc{nt}=0.44\,\mathrm{km\,s^{-1}}$. These dimensions are similar to those of SM1 in Oph-A region \citep{MAN98,ABMP07}. We alter the partition of kinetic energy in solenoidal modes (i.e. shear and rotation) and compressive modes (i.e. compression and rarefaction).

\subsection{Initial conditions}

The amplitude of a turbulent mode $\boldmath{a}(\boldsymbol{k})$ can be split into its longitudinal (compressive) component $\boldmath{a_\textsc{l}}(\boldsymbol{k})$ and transverse (solenoidal) component $\boldmath{a_\textsc{t}}(\boldsymbol{k})$ using Helmholtz decomposition:
\begin{equation}
  \begin{split}
    \boldsymbol{a_\textsc{l}}(\boldsymbol{k})&=\hat{\boldsymbol{k}}(\boldsymbol{a}(\boldsymbol{k})\cdot\hat{\boldsymbol{k}})\,,\\
    \boldsymbol{a_\textsc{t}}(\boldsymbol{k})&=\boldsymbol{a}(\boldsymbol{k})-\hat{\boldsymbol{k}}(\boldsymbol{a}(\boldsymbol{k})\cdot\hat{\boldsymbol{k}})\,.\\
  \end{split}
\end{equation}
When the magnitude and direction of the amplitude is random, there is on average twice as much energy in transverse modes as there is in solenoidal modes.

We define the parameter $\delta_\textsc{sol}$ as the average fraction of solenoidal kinetic energy in a velocity field. We modify the field to have given $\delta_\textsc{sol}$ by performing the transformation,
\begin{equation}
  \boldsymbol{a}(\boldsymbol{k})\to \sqrt{3\,(1-\delta_\textsc{sol})}\,\boldsymbol{a_\textsc{l}}(\boldsymbol{k})+\sqrt{\frac{3}{2}\,\delta_\textsc{sol}}\,\boldsymbol{a_\textsc{t}}(\boldsymbol{k})\,.
  \label{eqn:helm_trans}
\end{equation}

We generate ten initial cores, each with a unique random velocity seed. For each core, we apply the transformation in Eqn. \ref{eqn:helm_trans} with values $\delta_\textsc{sol}=0,\,\frac{1}{9},\,\frac{1}{3},\,\frac{2}{3},\,1$, yielding a total fifty core set-ups.

\subsection{Results}

\begin{figure*}
  \centering
  \subfigure[$\delta_\textsc{sol}=0$]{\label{mont_5}\includegraphics[height=0.33\textwidth]{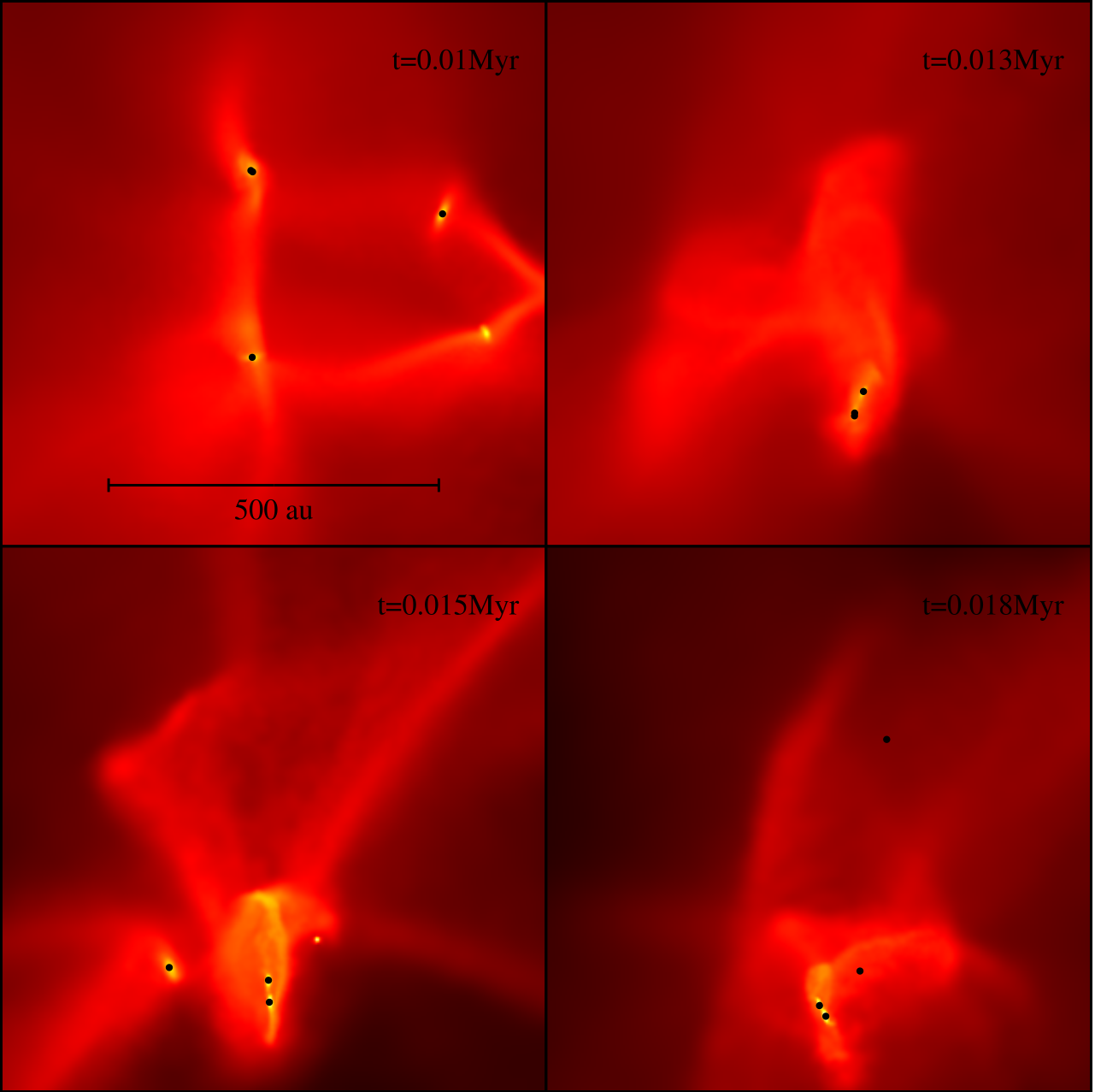}}
  \subfigure[$\delta_\textsc{sol}=1/9$]{\label{mont_4}\includegraphics[height=0.33\textwidth]{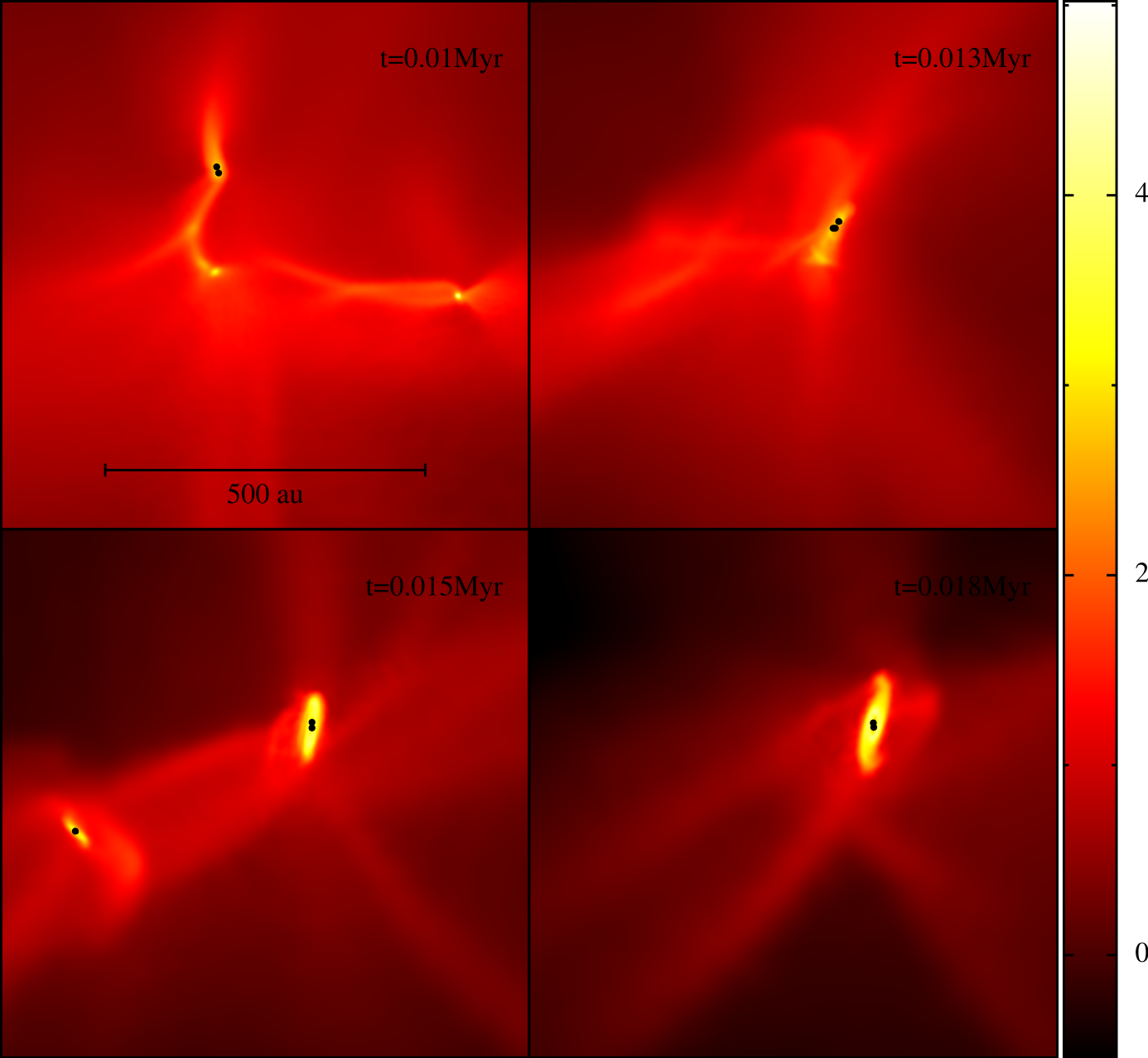}}
  \subfigure[$\delta_\textsc{sol}=1/3$]{\label{mont_3}\includegraphics[height=0.33\textwidth]{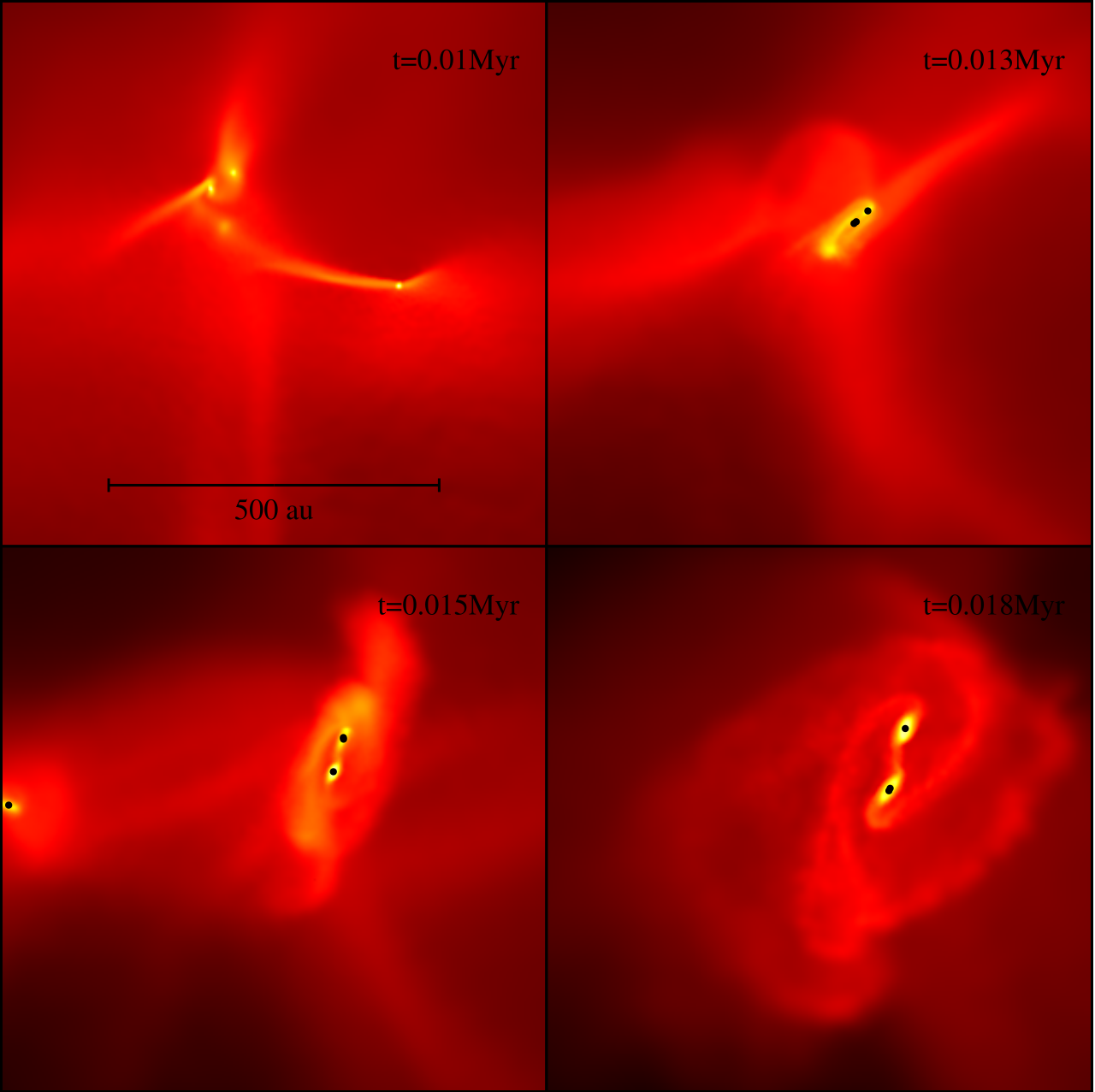}}
  \subfigure[$\delta_\textsc{sol}=2/3$]{\label{mont_2}\includegraphics[height=0.33\textwidth]{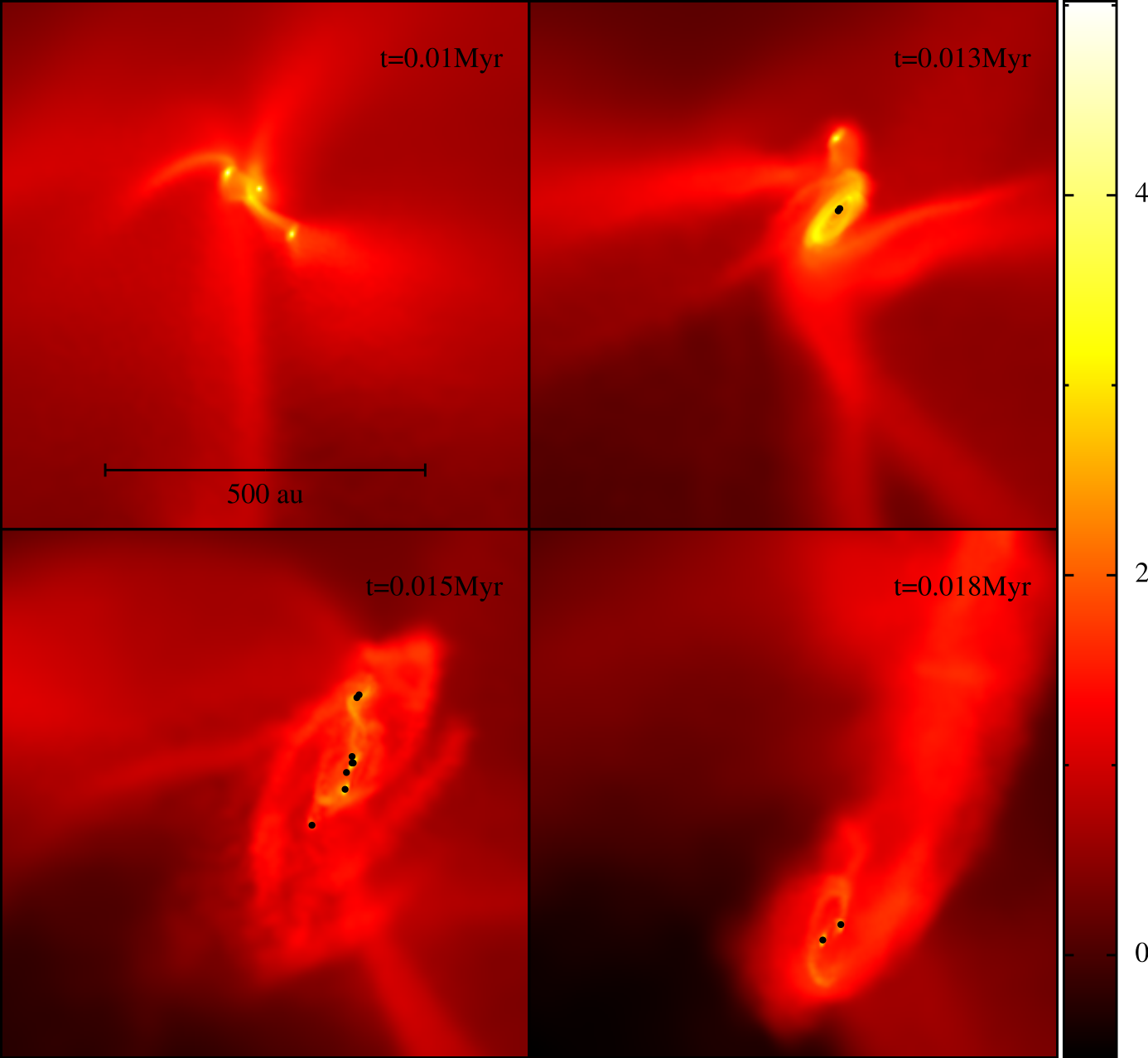}}
  \subfigure[$\delta_\textsc{sol}=1$]{\label{mont_1}\includegraphics[height=0.33\textwidth]{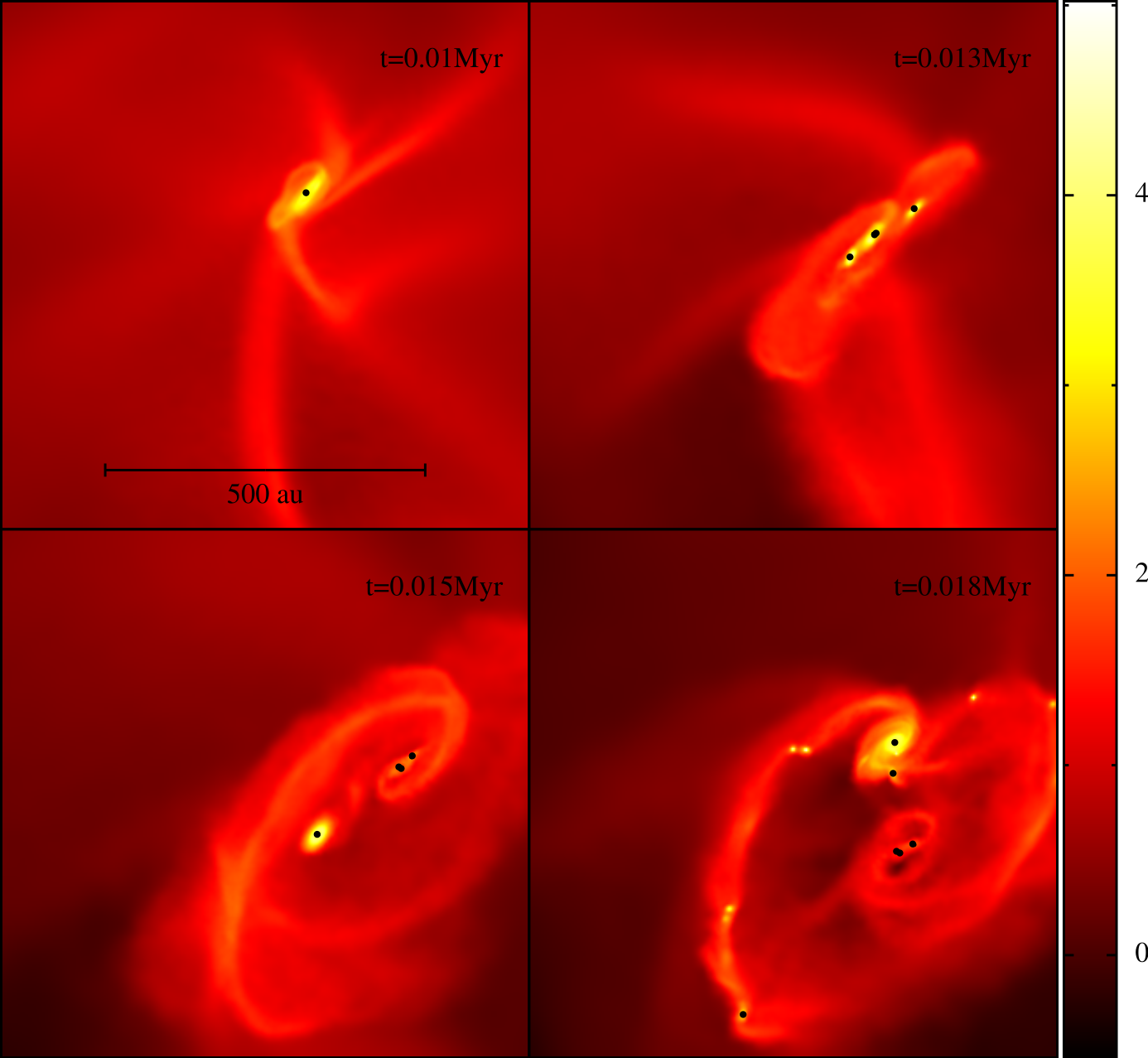}}
  \caption{Column density maps of the central $820\,{\rm au}$ by $820\,{\rm au}$ of the $(x,y)$-plane, from the simulations with fixed random seed and different values of $\delta_\textsc{sol}$, at times $t=1.00,1.25,1.50\text{ and }1.75\times10^4\,\mathrm{yrs}$. The colour scale gives the logarithmic column density in units of $\mathrm{g\,cm^{-2}}$. Sink particles are represented by black dots. \changes{ See \citet{LWH15} for the orginal version of this figure.}}
  \label{montage}
\end{figure*}

Fig. \ref{montage} shows a montage of simulation snapshots where the random seed is fixed and $\delta_\textsc{sol}$ is varied from 0 to 1\,. When the field is purely compressive ($\delta_\textsc{sol}=0$), protostars form within a network of filaments. Due to the low angular momentum of the system, the resultant protostars are only attended by small discs. In contrast, when the field is purely solenoidal ($\delta_\textsc{sol}=1)$, a single protostar forms and is attended by a coherent disc structure. The disc proceeds to fragment into multiple objects. A smooth transition between filament fragmentation and disc fragmentation is seen in the snapshots with intermediate values of $\delta_\textsc{sol}$.

\begin{figure}
  \includegraphics[width=\columnwidth]{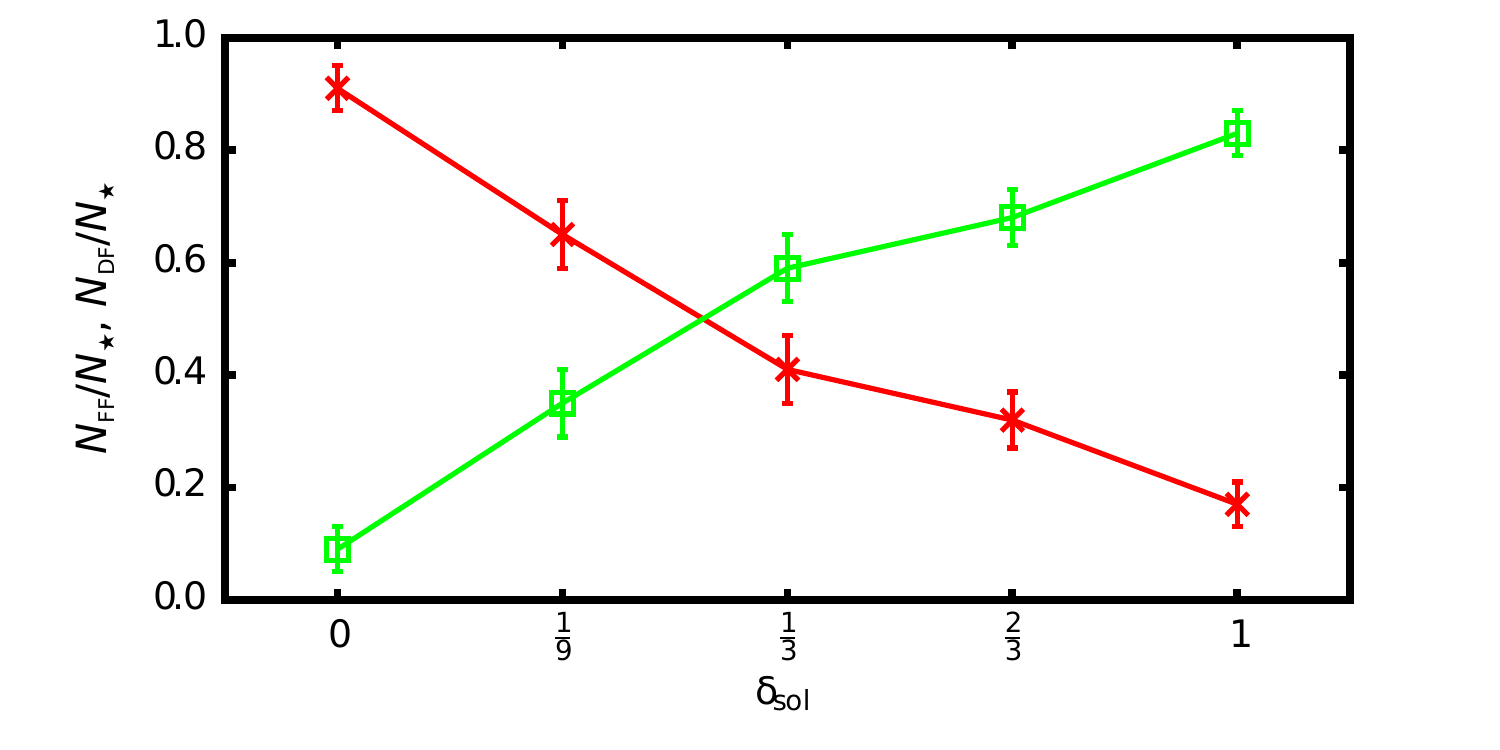}
  \caption{The fraction of stars formed by filament fragmentation (red crosses) and disc fragmentation (green boxes) for different values of $\delta_\textsc{sol}$, averaged over all random seeds. The error bars show the Poison counting uncertainties.\changes{ See \citet{LWH15} for the orginal version of this figure.}}
  \label{ff_df}
\end{figure}

Fig. \ref{ff_df} shows the fraction of protostars formed by filament fragmentation (i.e. in relative isolation) and disc fragmentation (i.e. in discs around more massive protostars) as a function of $\delta_\textsc{sol}$. These values are averaged over all random seeds. Here we see that the occurrence of disc fragmentation increases monotonically with $\delta_\textsc{sol}$. Filament fragmentation, therefore, decreases monotonically with $\delta_\textsc{sol}$.

\begin{figure}
  \includegraphics[width=\columnwidth]{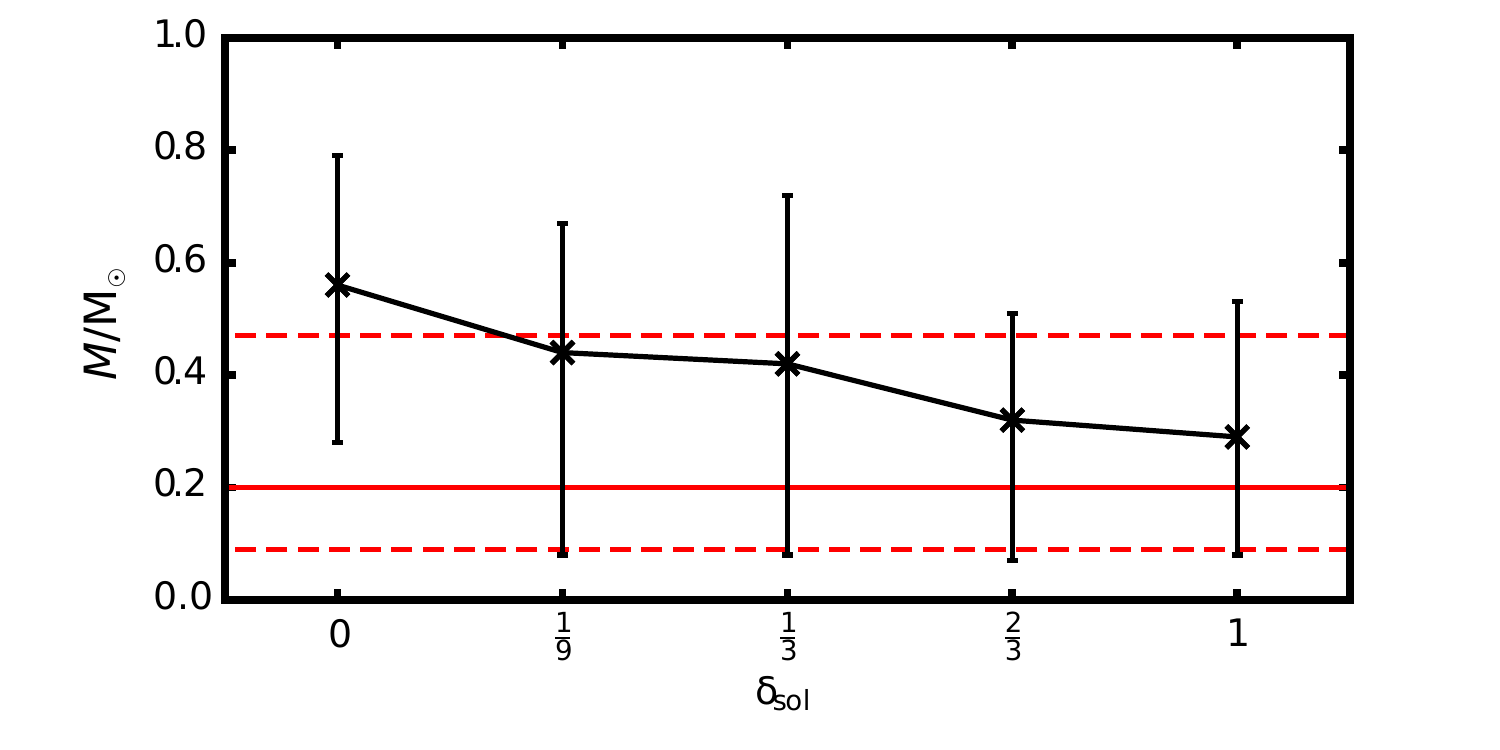}
  \caption{The black points show the median stellar mass, and the vertical black bars show interquartile range of mass, for different values of $\delta_\textsc{sol}$, averaged over all random seeds. The solid and dashed horizontal red lines show the median and interquartile range for the \citet{C05} IMF.\changes{ See \citet{LWH15} for the orginal version of this figure.}}
  \label{mass_median}
\end{figure}

On average, the number of protostars spawned per core ranges from $5.4\pm7$ when $\delta_\textsc{sol}=0$ to $8.1\pm9$ when $\delta_\textsc{sol}=1$\,. The median and interquartile range of mass is shown in Fig. \ref{mass_median}. Here we see that increasing $\delta_\textsc{sol}$ pushes the median sink mass down from roughly $0.6\,\mathrm{M_\odot}$ when $\delta_\textsc{sol}=0$ to $0.3\,\mathrm{M_\odot}$ when $\delta_\textsc{sol}=1$. We also find that purely compressive fields form very few brown dwarfs. We note that -- even for this limited set of initial conditions -- the interquartile range of protostellar masses when $\delta_\textsc{sol}\gtrsim2/3$ is very similar to that of the \citet{C05} IMF.

\subsection{Summary}

We show that the collapse and fragmentation of prestellar cores is strongly influenced by the structure of the velocity field. Disc formation and fragmentation dominates the star formation process when $\delta_\textsc{sol}\gtrsim1/3$. At values below this, stars form mostly through the fragmentation of filamentary structures. The value of $\delta_\textsc{sol}$ also affects distribution of protostellar masses. The distribution most resembles the observed IMF when $\delta_\textsc{sol}\gtrsim2/3$. Reducing $\delta_\textsc{sol}$ reduces the level of disc fragmentation, resulting in smaller number of objects with greater average mass. In extreme cases (i.e. $\delta_\textsc{sol}<1/9$), the formation of low mass stars and brown dwarfs is heavily suppressed. This suggests that disc fragmentation may be a requirement for forming these objects.

\section{SUMMARY AND CONCLUSIONS}
\label{sec:summary}

Previous numerical work \citep[e.g.][]{SHW07,SW09a} shows that disc fragmentation is important mechanism for reproducing the properties of low mass stars. We demonstrate that disc fragmentation also plays an important role in the conversion of prestellar cores into stars. Importantly, the observed masses and multiplicities of stars can be recovered from simulations if the following criteria are satisfied:
\begin{itemize}
 \item \emph{Radiative feedback from accretion onto protostars is episodic.} Simulations with episodic radiative feedback produce both an IMF and multiplicity statistics in good agreement with those observed. Simulations with continuous radiative feedback fail to produce both the observed number of brown dwarfs and the multiplicity statistics associated with young objects. Furthermore, continuous radiative feedback produces protostellar luminosities greater than those observed in young stars. Simulations with no radiative feedback can produce good multiplicity statistics and a reasonable -- albeit bottom-heavy -- IMF, but are unrealistic.
 \item \emph{A significant proportion of the core's internal kinetic energy is in solenoidal turbulent modes.} Cores with more than a third of their kinetic energy in solenoidal modes are able to easily produce stars via disc fragmentation. Decreasing this fraction results in more stars forming in fragmenting filaments. These objects tend of be of greater mass, resulting in a top-heavy mass distribution relative to the observed IMF. Furthermore, filament fragmentation struggles to produce brown dwarfs and very low mass stars.
\end{itemize}

\section*{Acknowledgements}

OL and APW gratefully acknowledge the support of a consolidated grant (ST/K00926/1) from the UK STFC. This work was performed using the computational facilities of the Advanced Research Computing @ Cardiff (ARCCA) Division, Cardiff University. All false-colour images have been rendered with \textsc{splash} \citep{P07}.
 
\bibliographystyle{apj}
\bibliography{refs}
\end{document}